\RequirePackage{lineno}
\documentclass[twocolumn]{revtex4}

\usepackage{amsmath,amsfonts,amssymb}
\usepackage{graphicx}
\usepackage{natbib}
\usepackage{placeins}  % Add in the preamble
\setcitestyle{authoryear,round}
\usepackage{lineno}  
\begin{document}

% \title{On the dynamic feasibility of mammalian interactions}
\title{Coarse-Graining Cascades Within Food Webs}

\author{%%%% Author details
Justin D. Yeakel$^{1,2,*}$}

%%%%%%%%% Insert author address here
\address{
$^{1}$Life \& Environmental Sciences, University of California Merced, Merced, California, USA\\
$^{2}$The Santa Fe Institute, Santa Fe, New Mexico, USA\\
$^{*}$Corresponding author: jyeakel@ucmerced.edu}

% \textit{Keywords}: Consumer-resource interactions, food webs, state dynamics, trophic cascades.

%150 words
%TC:ignore
\begin{abstract}

Quantifying population dynamics is a fundamental challenge in ecology and evolutionary biology, particularly for species that are cryptic, microscopic, or extinct. Traditional approaches rely on continuous representations of population size, but in many cases, the precise number of individuals is unknowable. Here, we present a coarse-grained population model that simplifies population dynamics to binary states -— high or low -— determined by the balance of bottom-up resource availability and top-down predation pressure. This Boolean framework provides a minimal yet analytically tractable alternative to traditional Lotka-Volterra-based models, enabling direct insights into the role of food web structure in shaping community stability. Using this approach, we investigate how trophic interactions influence population persistence, cyclic dynamics, and extinction risk across model food webs. We find that top-down effects are a primary driver of cycling, aligning with theoretical expectations from traditional population models, and that trophic position strongly influences extinction risk, with higher-trophic species more prone to persistent low-population states. Additionally, we explore the role of trophic short-circuits -- direct interactions between apex predators and low-trophic prey -- and find that they can buffer cascades and alter extinction patterns in ways that are often overlooked in classical models. By simplifying population dynamics to a two-state system, this framework provides a powerful tool for disentangling the structural drivers of community stability. These results highlight the potential of coarse-grained approaches to complement existing models, offering new insights into trophic interactions, extinction risks, and the susceptibility of species to trophic cascades.
\end{abstract}
% \newpage{}

\maketitle

\section*{Introduction}

%THIS NEEDS TO BE SHORT

%From numbers to state dynamics
Quantifying the flux of individuals into and out of populations is a central challenge that lies at the intersection of ecology and evolutionary biology.
In a real sense, a population's true density, and how it changes over time, will be known only in small or artificial environments, with the remaining task to grasp the uncertainties in both the data available and the models used to infer population trends. \citep{oppel2014assessing}.
In many cases -- especially for cryptic, microscopic, or extinct species -- documenting population dynamics is either infeasible or impossible.
Frameworks designed to investigate changes in population state, rather than changes in the numbers of indidviduals or density, may more accurately represent our potential knowledge of a system, and by their simplicity, present an opportunity for disentangling the role of interaction structure on community dynamics.

%Importance of state dynamics of populations in ecology
One typically cares only about numbers to the extent that they communicate changes in state.
At its simplest, states associated with organismal populations can be categorized as small (high risk) and large (low risk).
This perspective effectively coarse-grains a traditional population dynamic, where an arbitrary threshold delineates whether the population occupies a low versus high state \citep{robeva2016spruce,guimaraes2024dynamics}.
Coarse-graining population dynamics to a small set of states has been a common approach in metapopulation theory, popularized by the core-satellite species concept where some species occupies many patches with high occupancy (core), while others persist in fewer patches with low occupancy (satellite) \citep{hanski1982dynamics}.
\citet{hanski1985single} integrated these concepts into a dynamic framework where low and high population states were directly used to explore metapopulation fragmentation, rescue, and collapse.
More recent efforts to compare the dynamical outcomes of such two-state Boolean population models against traditional ODE approaches point to strong alignment in stability regimes \citep[e.g. in the spruce-budworm model;][]{robeva2016spruce}.
It is thus likely that formal links connecting low-dimensional Boolean frameworks to higher-dimensional and more commonly used population models exist, but are largely unexplored, particularly in ecological contexts.

%To networks
The utility of minimal dynamical frameworks may best be realized in large systems of interacting components, such as food web networks where nodes represent species, and links represent the trophic interactions connecting them.
Here, structural and dynamical complexity intersect, often clouding our undestanding of what mechanisms drive which outcomes.
Boolean frameworks are widely used to explore large systems with many interactions, such as those developed to simulate gene regulatory networks \citep{Kauffman1971}.
Related approaches integrating low-dimensional state dynamics have been applied to ecological networks to examine ecosystem services \citep{ross2021universal,guimaraes2024dynamics}, disease epidemics \citep{connolly2022epidemic}, community assembly \citep{campbell2011network,yeakel2020diverse}, and metacommunity dynamics \citep{leibold2004metacommunity,pillai2010patch,gross2020modern}.
% , and even used to untangle empirical time series.

%The problem we are targeting
Integration of low-dimensional Boolean frameworks to understand the role of interactions within food webs has been limited, in part due to the fact that it is difficult to generalize acoss sytems with alternative algorithmic rules \citep{Seshadhri2016}.
Yet exploring the dynamics of complex ecological networks when dynamical rules are minimal may allow more direct insight into the role of interaction structure, and in particular shed light on the interplay between top-down and bottom-up forcing on species populations within and among food webs.

%What we are doing 
Here a coarse-grained population framework is introduced, where species' populations interacting within a food web occupy a high or low state, depending on the intersection of bottom-up forcing from their resources and top-down forcing from their predators.
The architecture of this approach is effectively Boolean in nature, where if bottom-up influences promoting growth are much larger than the top-down influences promoting decline, a species' population is more likely to occupy a high rather than low state, and vice versa.
This framework is leveraged to explore the contrasting forces at work within model food webs, as one moves from the lowest to highest trophic levels, and is specifically applied to understand the potential for trophic cascades as a function of species position within the system.
The coarse-grained model described here is tractable to both numerical and analytical exploration, where dynamical outcomes are shown to align with theoretical and empirical predictions associated with population cycles, trophic-level specific extinction risk, and the potential for disturbances to cascade across interacting species.
Of particular interest is the role of trophic short-circuits, where apex predators interact with lower-trophic species, a subject that has recieved little attention, but, for example, may be essential for understanding the historical roles of mysticete whales in marine food webs.
Taken together, initial exploration of this coarse-grained population model suggests its tractability and low-dimensionality may be useful for understanding how interaction structures shape community dynamics and the emergence of trophic cascades.

\section*{Model Description}

%Intro and roadmap
Here the dynamics of populations are coarse-grained to capture a small number of changing states, rather than changing numbers or densities of individuals.
So rather than a species $i$'s population being described by a continuous or discrete range of values, it is instead coarsened to state $S_i$, where $S_i = S^+ \equiv +1$ represents a relatively large population, and $S_i = S^- \equiv -1$ represents a relatively small population.
(The numerical values of each state are not important. States of $S_i = \pm 1$ are chosen for simplicity and ease of computation.)
What influences a small population to become large, or a large population to become small is described by the state dynamics, which follow a coarse-graining of traditional equations that define changes in populations over discrete time intervals.
% This contribution specifically focuses on 
Here it is the influence of trophic interactions between species in a food web that determines whether and in which direction a population's state is expected to change.
Below the dynamics of population state are described, followed by the procedure used to generate model food webs.

\textbf{Dynamics.} 
A population occupies one of two states at time $t$ -- low $S_i(t) = S^-$ or high $S_i(t) = S^+$ -- where the forces that compel a state change are assumed to be trophic in nature.
% Throughout, $S^\pm = \pm 1$ such that zero is the demarkation between states, though this choice is arbitrary.
As such, the bottom-up consumption of $n_i$ resources by species $i$ at time $t$ will promote $S_i(t+1) = S_i^+$, whereas top-down mortality from $m_i$ predators will promote $S_i(t+1) = S_i^-$.
Primary productivity is captured by $g$, which is a constant influx to all primary producers occupying the lowest trophic level.
The magnitude of these bottom-up and top-down effects is denoted by the parameters $k^{\rm in}$ and $k^{\rm out}$, respectively, both of which are assumed to be constant across the food web.
Taken together, the state dynamics of the coarse-grained system is given by
% \begin{equation}
%     S_i(t+1) = {\rm sign} \left[ S_i(t) + k^{\rm in}\sum_{j=1}^{n_i}S_j(t) + \theta_i g - k^{\rm out}\sum_{l=1}^{m_i}S_l(t) \right],
%     \label{eq:statedynamics}
% \end{equation}
\begin{equation}
    F(S_i(t)) = S_i(t) + k^{\rm in} \sum\limits_{j=1}^{n_i} S_j(t) + \theta_i g - k^{\rm out} \sum\limits_{l=1}^{m_i} S_l(t),
    \label{eq:statefunction}
\end{equation}
where
\begin{equation}
    S_i(t+1) =
    \begin{cases} 
        +1, & \text{if } F(S_i(t)) > 0, \\
        -1, & \text{if } F(S_i(t)) < 0, \\
        S_i(t), & \text{if } F(S_i(t)) = 0,
    \end{cases}
    \label{eq:statedynamics}
\end{equation}
assuming $\theta_i = 1$ for all primary producers and zero otherwise.

%Structure
\textbf{Structure.} Food web construction is implemented using the niche model, described in detail in \citet{williams2000simple}, and analytically investigated in \citet{camacho2002analytical,stouffer2005quantitative}.
The general procedure for generating food webs is as follows.
First, each of $S$ species is assigned a niche value, $\eta_i$, drawn uniformly from the niche interval $[0,1]$, and assumed to prey on all species within a feeding range $r_i$.
The range is defined as $r_i = x_i \eta_i$, where $x_i \sim {\rm Beta}(a,b)$, given $a = 1$ and $b = (2C)^{-1} - 1$, and where $C$ is the connectance of the food web.
Connectance is defined as $C = L/S(S-1)$, with $L$ representing the number of links in the food web network. 
The center of the feeding range, $c_i$, is drawn uniformly from the interval $[r_i/2, \eta_i]$, thus allowing for looping and cannibalism. 
Using these ranges, an adjacency matrix is constructed to encode trophic interactions, and species without any prey or predators are iteratively removed. 
This ensures the resulting food web is connected, with all species participating in at least one trophic interaction. 

The niche value assigned to each species $\eta_i$ is a rough proxy of trophic level $\tau_i$.
Species with low niche values are more likely to be primary producers; species with high niche values are more likely to be apex predators.
Despite this qualitative connection, it is difficult to extrapolate trophic level from niche value given the iterative stacking of food webs from randomly generated feeding interavals.
In order to convert niche values assigned to species to their expected trophic level (where the utility of this mapping is described in more detail below), a statistical relationship, fitted across food web replicates, is instead employed, where 
\begin{equation}
\tau = 1 + z_0 \eta^{z_1}, 
\label{eq:TL}
\end{equation}
where $(z_0,z_1)$ are fitted parameters.
This relationship changes with food web connectance $C$, such that more connected food webs tend to have elevated trophic levels for a given niche value (see Supplementary Fig. \ref{fig:tlniche}).

\section*{Results}

%Intro and roadmap
The state dynamics of populations in food webs showcase a range of temporal patterns, from steady states (consistently low or high population states) to cycles with alternating durations in each state.
Simplified state dynamics allow for both simulation and analytical approximation of food web dynamics, facilitating an understanding of the forces driving these patterns.
A general roadmap to this initial investigation is as follows.
First, the conditions giving rise to population cycles is described, and the dependency of state dynamics on bottom-up or top-down pressures, primary productivity, and trophic level is investigated.
Second, the probability that a population exists in a low state, or $p(S^-)$, is explored by integrating an analytical approximation with simulation results, providing insight into the role of asymmetric interaction pressures on state dynamics.
Third, traditional measures of trophic cascades are mapped to the course-grained framework, and the relationship between cascade magnitude, trophic level, and bottom-up versus top-down forcing is evaluated both analytically and numerically.

% and both the prevalance and drivers of cascades is explored.

\subsection*{State dynamics change with top-down versus bottom-up influence}

\begin{figure}[ht]
    \centering
    % First panel
    % \begin{subfigure}[b]{0.45\linewidth}
    %     \centering
    %     \includegraphics[width=\linewidth]{fig_web.pdf}
    %     % \caption{Food web in final state.}
    %     \label{fig:web}
    % \end{subfigure}
    % \hspace{0.05\linewidth} % Horizontal space between panels
    % % Second panel
    % \begin{subfigure}[b]{0.45\linewidth}
    %     \centering
    %     \includegraphics[width=\linewidth]{fig_sim.pdf}
    %     % \caption{Proportion oscillating.}
    %     \label{fig:sim}
    % \end{subfigure}
    \includegraphics[width=0.95\linewidth]{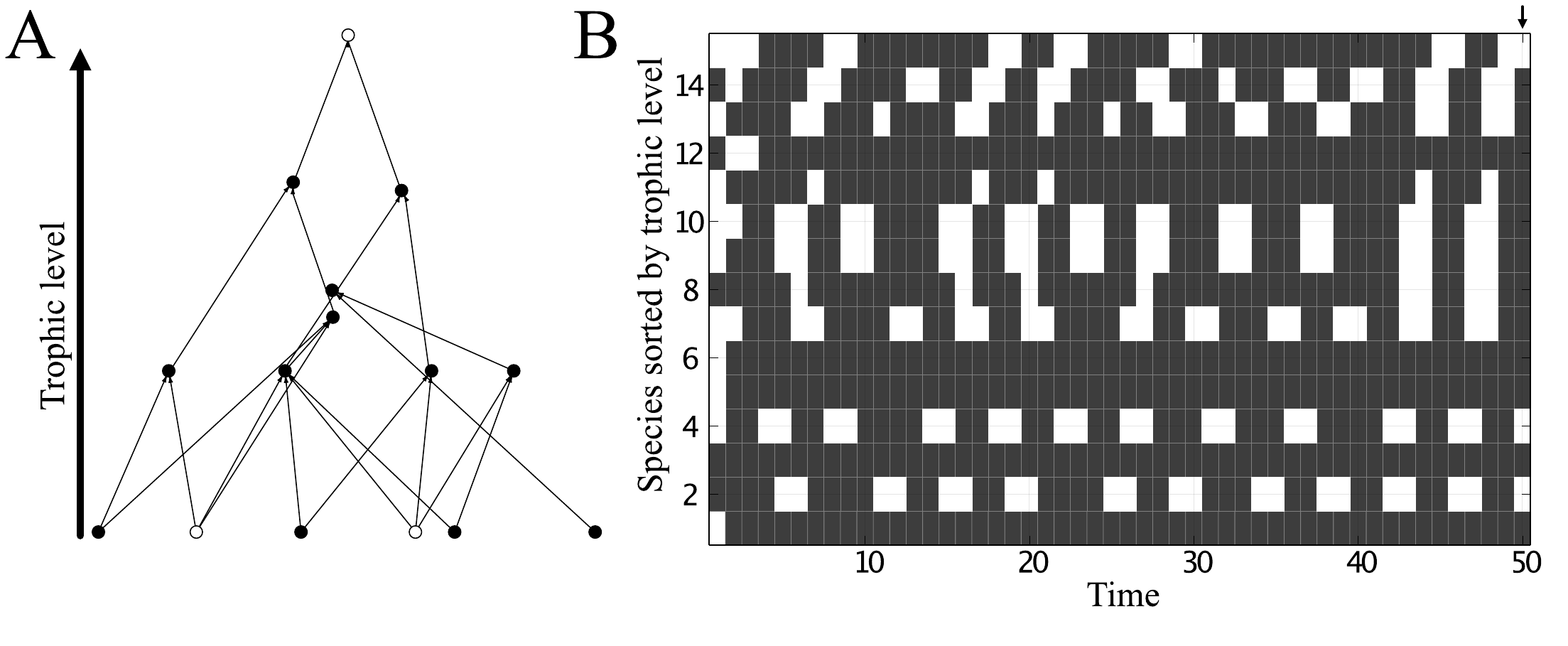}
    \caption{
        A. An example food web, where black nodes denote species in high population states, and white nodes depict species in low population states, here documenting the system state at the terminal time $t = t_{\rm max}$.
        Species are depicted from the lowest (bottom) to highest (top) trophic levels.
        B. A temporal sequence of state dynamics, where time is on the x-axis and species are arrayed on the y-axis from the lowest to highest trophic level.
    }
    \label{fig:foodweb}
\end{figure}

%Example of output
An example of a small food web and the resulting state dynamics is shown in Fig. \ref{fig:foodweb}. %\citep[see][]{Habermann}.
From a random initial state, populations switch from low (white) to high (gray) populations and back, depending on \emph{i}) the influence of primary productivity $g$, the contribution of bottom-up pressures -- driven by $k^{\rm in}$ and the number of resources allocated to species $i$, $n_i$, and the contribution of top-down pressures -- driven by $k^{\rm out}$ and the number of predators impacting species $i$, $m_i$.
In the dynamics shown in Fig.\ref{fig:foodweb}B, some species achieve steady state dynamics, whereas others cycle.
Some cycles have a regular period; others have more irregular or complex periods.
At the lowest trophic level, those populations with more than two predators appear to cycle, whereas those with fewer do not.
In this example, primary productivity from basal resources $g=5$ and the magnitude of top-down effects $k^{\rm out} = 2$.
As such, when there are fewer than three predators, the top-down forces are overcome by the constant supply of basal resources and those lowest trophic species remain at high population states.
When there are three or more predators, primary producers may cycle, depending on the combined population states of their predators.
At higher trophic levels, such a simple dichotomy breaks down, as the combinations of bottom-up and top-down forces become more complex.

\begin{figure}[ht]
    \centering
    \includegraphics[width=0.95\linewidth]{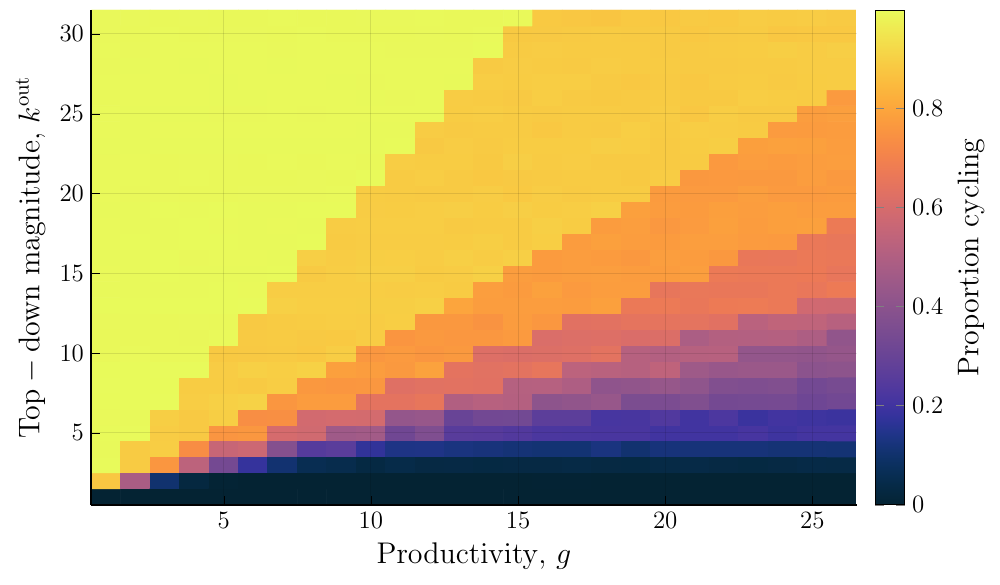}
    \caption{The proportion of cycling species within food webs as a function of primary productivity, $g$, and the magnitude of top-down effects, $k^{\rm out}$. Values are averaged across 200 replicate food webs, where $k^{\rm in} = 5$, $t_{\rm max} = 300$ (discarding the first 200 time steps) and under the assumption of random initial conditions.
    }
    \label{fig:propcycling}
\end{figure}

%Macroscale - probability of cycling
While the cycling of individual populations depends on specific interactions with resources and predators, the potential for cycling writ large is observed to be driven by both the primary productivity $g$, as well as the magnitude of top-down forces $k^{\rm out}$ relative to bottom up forces $k^{\rm in}$.
Simulation results reveal that the proportion of populations cycling within food webs increases with primary productivity, as long as the magnitude of top-down froces is on par with, or exceeds, the magnitude of bottom-up forces (Fig. \ref{fig:propcycling}).
Similarly, the proportion of populations that cycle also increases with the magnitude of top-down forces, though this effect is more gradual when primary productivity is high.
In cases where the top-down forces are extreme, and primary productivity is low, nearly all of the populations in a given food web are prone to cyclic dynamics.

While the potential for cycling between states is often considered a measure of population instability \citep{Krebs1996}, it can simultaneously be percieved as a measure of resilience, given that what is low in a cycling population will soon recover.
Of more interest, perhaps, is the temporal duration of populations in different states as they cycle.
The duration each population spends in low $S^-$ and high $S^+$ states follows a similar pattern as the proportion of populations that cycle, but showcases two competing effects.
First, the average duration of populations in a high state $S^+$ increases as the magnitude of top-down effects decrease, and as primary productivity $g$ increases.
In words, when bottom-up forces are strong, populations -- on average -- tend to remain in a high state for up to 100\% of simulation time.
Second, the reverse is true for populations in a low state $S^-$, though the average duration is constrained to fall between 0\% and 50\% of simulation time.
In words, when top-down forces outweigh bottom-up effects, populations maximize their time in low states, but this occurs in cycling regions where each fall to $S^-$ is matched by a subsequent recovery to $S^+$.
While these patterns clearly correlate across top-down versus bottom-up pressures, they represent averages across the food web writ large, and potentially mask realized dynamics between interacting species within a given food web, which may vary across trophic levels.

\begin{figure*}[ht]
    \centering
    \includegraphics[width=0.95\linewidth]{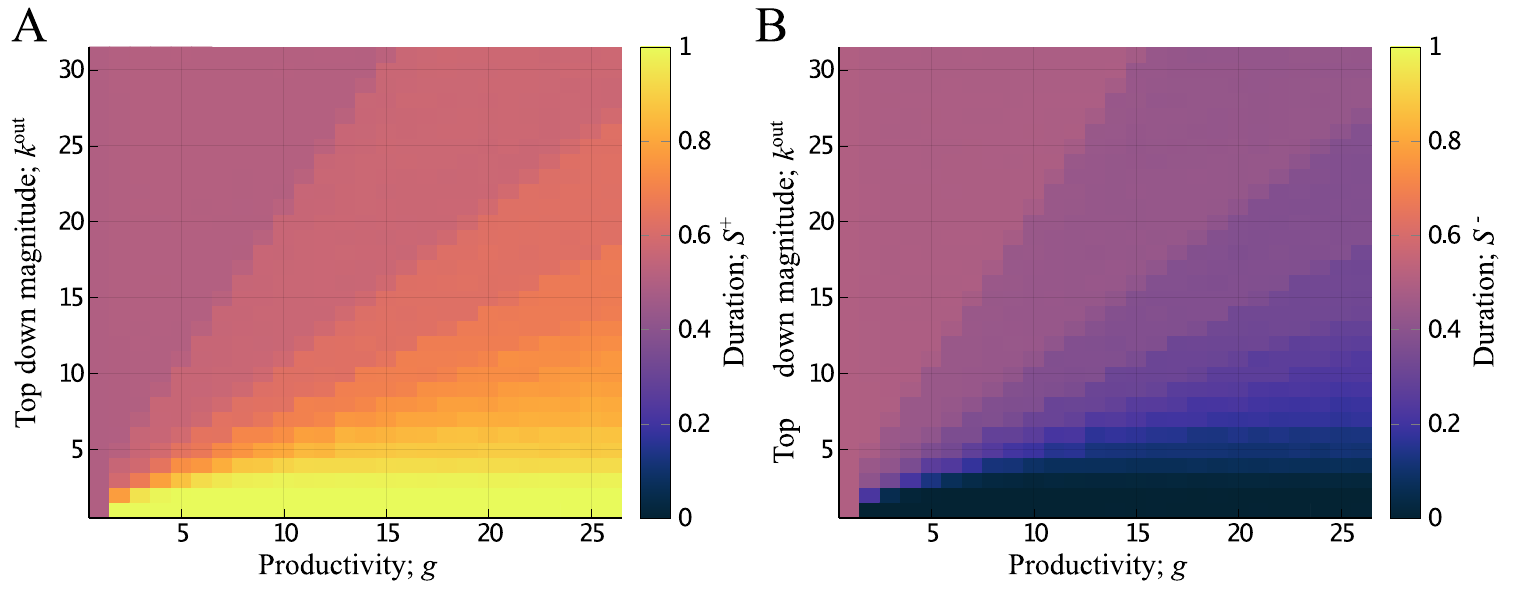}
    \caption{
        Temporal duration measured as the proportion of simulation time occupied by A. the high population state $S^+$ and B. the low population state $S^-$, as a function of primary productivity $g$ and the magnitude of top-down forces $k^{\rm out}$. Values are averaged across 200 replicate food webs, where $k^{\rm in} = 5$, $t_{\rm max} = 300$ (discarding the first 200 time steps) and under the assumption of random initial conditions.
    }
    \label{fig:duration}
\end{figure*}

%Duration as a function of trophic level
To understand the effects of trophic level on state dynamics, simulations with the particular conditions of $k^{\rm out} = 10$, $k^{\rm in} = 5$, and $g = 10$, are explored in detail.
Here, it is observed that the duration of low verses high states changes with trophic level.
% In other words, the food web dynamics that are considered exist within an region where the proportion of cycling populations is intermediate and between extremes (Fig. \ref{fig:propcycling}).
% Here the populations of primary producers at the lowest trophic level are fed by an external subsidy $g$, which represents the primary productivity of the system.
At the bottom of the food web, where species are fed by the basal resource $g$, there is a strong tendancy for this species to exist in a high population state (blue line, Fig. \ref{fig:duration}).
% Here, increased $g$ will strongly bias lower-trophic species towards high population states (Fig. \ref{fig:durationtrophic}), though excessive predation can still have a negative impact on these species.
For those species approaching trophic level 2, the duration of species in high population states declines, increasing slightly for those approaching trophic level 3, and declining again for species at the top of the food web.
% , though this region is also characterized by increased variability (CHECK).

\begin{figure}[ht]
    \centering
    \includegraphics[width=0.95\linewidth]{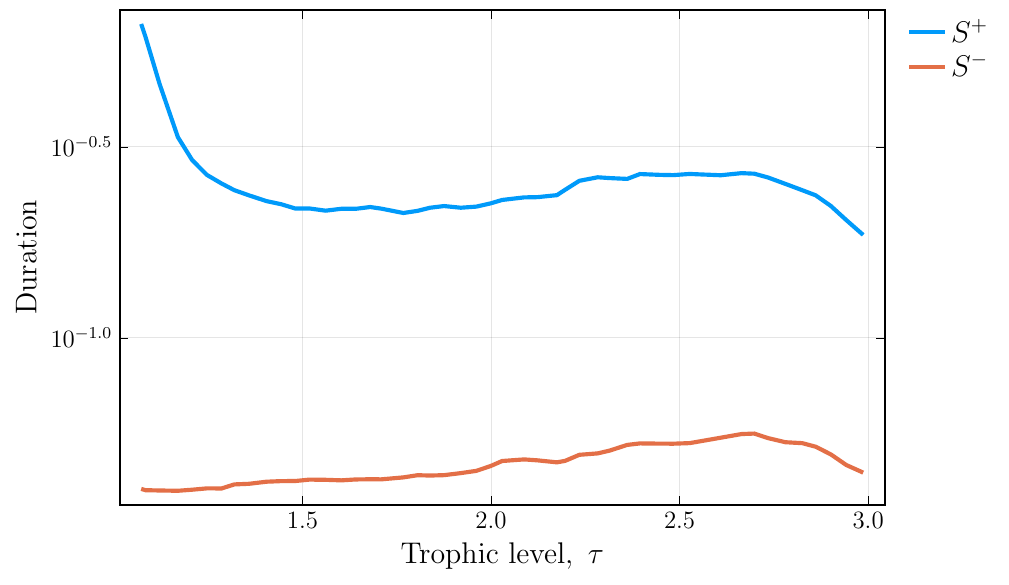}
    \caption{Duration as the proportion of simulation time occupied by a high population state $S^+$ (blue line) and low population state $S^-$ (red line) as a function of trophic level.
    Parameters were set as $S=100$, $C=0.02$, $k^{\rm in} = 5$, $k^{\rm out} = 10$, $g=10$, and duration was averaged across 5000 replicate food webs.
    With a simulation time $t_{\rm max} = 300$, the first 200 time steps were discarded to avoid transient effects.}
    \label{fig:durationtrophic}
\end{figure}

In contrast, the duration of populations in a low state represents a much smaller percentage of simulation time, with a very low duration for primary producers (again due to the influence of $g$), increasing as populations approach trophic level 3, and declining slightly at the highest trophic level (red line, Fig. \ref{fig:duration}).
That the apex predators would rarely occupy the low population state is not surprising, as these populations have only bottom-up effects that are left un-countered by top-down regulation.
While consumption of prey that are themselves in low population states may not be enough to maintain $S^+$, the likelihood is still largely biased in favor of a high population state.
It is this asymmetry in bottom-up versus top-down forces -- due to trends in the numbers of predators and prey as one moves the bottom to the top of the food web -- that will enable some analytical approximation of state dynamics to be investigated, which is explored next.

%Analytical. Motivate with p(S-)
\subsection*{Drivers of state dynamics across trophic levels}

The concepts of cycling between states and duration within states can in part be captured by the probability that a particular population achieves a particular state over time.
For example, if the probability that a population is in a low state $p(S^-) = \{0, 1\}$, the probability of cycling will be zero, and duration will be unity.
Alternatively, if $0 < p(S^-) < 1$, the probability of cycling is unity, whereas that of duration will be less than unity.
As $p(S^-)\rightarrow 0.5$, duration will follow in a similar fashion.
It is thus desirable to explore an analytical expression of $p(S^-)$, and because state duration clearly varies with trophic level, $\tau$, understand how the bottom-up and top-down forces imposed by species interactions contributes to changes in $p(S^-)$ as one moves from the bottom to the top of the food web.

%Transition to analytical exploration of prob(S-)
To build an approximate solution for $p(S^-|\tau)$, a sigmoidal function is employed to capture the change in state as a function of the balance (or lack thereof) between top-down versus bottom-up forces.
From Eq. \ref{eq:statedynamics}, we can delineate the change in state due to bottom-up effects on species $i$ as
% \begin{equation}
    $\Delta_i^{\rm BU} = k^{\rm in}\sum_{j=1}^{n_i}S_j(t) + \theta_i g,$
    % \label{eq:bu}
% \end{equation}
and that due to top-down effects as
% \begin{equation}
    $\Delta_i^{\rm TD} = k^{\rm out}\sum_{l=1}^{m_i}S_l(t)$.
    % \label{eq:td}
% \end{equation}
Thus, the probability that a species $i$ occupies a low population state can be written
\begin{equation}
    p(S_i^-) = 1 - \frac{1}{1 + {\rm e}^{-(\Delta_i^{\rm BU}-\Delta_i^{\rm TD})}}.
    \label{eq:pS}
\end{equation}
The expectations of bottom-up and top-down pressures can then be expressed as 
\begin{align}
{\rm E}\{\Delta_i^{\rm BU}\} &= k^{\rm in} {\rm E}\{n_i\}(1 - 2 p(S_{\rm res}^-)),~{\rm and} \nonumber \\
{\rm E}\{\Delta_i^{\rm TD}\} &= k^{\rm out} {\rm E}\{m_i\}(1 - 2 p(S_{\rm pred}^-)),
\end{align}
respectively.

% What remains is to understand how the factors influencing $\Delta^{\rm BU}$ and $\Delta^{\rm TD}$ change as we move from the bottom to the top of the food web.

The use of the niche model to randomly construct food webs with structures similar to those of observed communities provides some basic tools that will be useful for establishing an expectation of $p(S_i^-|\eta)$, which can be subsequently converted to $p(S_i^-|\tau)$ using Eq. \ref{eq:TL}.
First, the expected number of resources and predators is observed to change in predictable ways from low to high niche values, and by extention, trophic levels.
% (approximating low to high trophic levels, a connection that will be explored further below).
Applying the simplifying assumption that the prey range of species $i$ is strictly below its niche value $\eta_i$, we obtain an approximation for the expected number of resources as a function of niche value
% \begin{equation}
    ${\rm E }\{n_i\} \approx 2(S-1)C\eta_i$.
% \end{equation}
Similarly, we can approximate the expected number of predators as a function of niche value as
% \begin{equation}
   $ {\rm E}\{m_i\} \approx 2(S-1)C(1-\eta_i)$.
% \end{equation}
See Supplementary Fig. \ref{fig:predprey} for a comparison of these approximations against model webs.
Two approaches are used to construct expectations for $\Delta_i^{\rm TD}$ and $\Delta_i^{\rm BU}$:
\emph{i}) resource state probabilities $p(S_{\rm res}^-)$ are calculated recursively from primary producers upwards along the niche axis, while
\emph{ii}) predator state probabilities are assumed to be, on average, constant, such that $p(S_{\rm pred}^-)=c_0$. % evaluated to understand the influence of alternative top-down forcings.

\begin{figure*}[ht]
    \centering
    \includegraphics[width=0.95\linewidth]{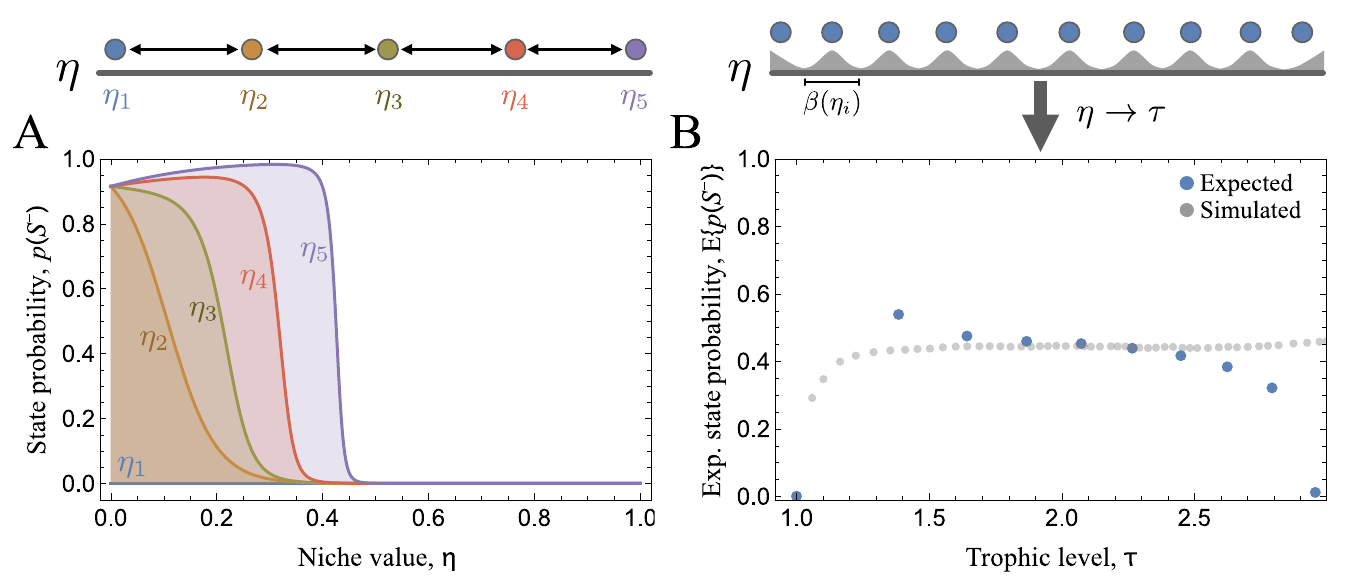}
    \caption{
        A. The probability of a low population state $p(S^-|\eta_i)$ for a set of rank-ordered species along the niche interval, as a function of each species $i$'s niche value $\eta_i$.
        B. The expected probability of a low population state ${\rm E}\{p(S_i^-)\}$ across a range of potential niche values, for a set of rank-ordered species along the niche interval, where niche values have been converted to trophic level (Eq. \ref{eq:TL}). 
        Simulation results (gray line) are shown for comparison. 
    }
    \label{fig:probS}
\end{figure*}

The influence of resources on $p(S_i^-)$ is calculated recursively, starting with the primary producers, and moving up along the niche interval.
Given a rank-ordered set of species with niche values constrained to the interval $\eta_i\in(0,1)$ (see top graphic in Fig. \ref{fig:probS}A), we can write
\begin{equation}
    {\rm E}\{\Delta_i^{\rm BU}\} = k^{\rm in}2(S-1)C\eta_i(1 - 2 {\rm E}_{\rm res}\{p(S^-)\}).
\end{equation}
Here the expectation ${\rm E}_{\rm res}\{p(S^-)\}$ is taken across all potential resource species with niche values $\eta_j<\eta_i$.
% For primary producers, ${\rm E}\{p(S_i^-)\} \approx 0$ as long as primary productivity $g$ is large enough to outweigh the effects of potential top-down pressures. 
In contrast, the influence of top-down predator states is assumed to be constant, such that $p(S_{\rm pred}^-)=c_0$, and where insight can be gained by observing how changes in $c_0$ in turn alters the expected state of species $i$.
Simulation results suggest that, on average, higher-trophic predator states $p(S_{\rm pred}^-)$ are slightly less than 1/2, and unless otherwise specified it is assumed $p(S_{\rm pred}^-)=0.47$ throughout.
% by assuming different relationships for $p(S_{\rm pred}^-)=f(\eta)$ and observing consequent impacts on state probabilities, a clearer understanding of top-down effects can be gained.
% The simplest mean-field assumption is that top-down predator state probabilities are constant across the niche axis, such that $f(\eta) = c_0$.
% Alternatively, simulation results suggest that there is a slight increase in $p(S^-)$ along the niche axis, though values do not stray far from 0.5 in either direction.
% In this case, predator state probabilities can be assumed to increase linearly from 0.45 to 0.55 as one ascends the niche axis.
After accounting for ${\rm E}\{\Delta_i^{\rm TD}\}$ and ${\rm E}\{\Delta_i^{\rm BU}\}$, $p(S_i^-|\eta_i)$ can then be approximated as a function of the species $i$'s actualized niche value $\eta_i$ (Fig. \ref{fig:probS}A), which may vary along the niche axis though the rank order is assumed to remain unchanged.

% Evaluating $p(S_i^-|\eta_i)$ for species as a function of their niche value $\eta_i$, a distribution of values along the niche axis is obtained.
When species $1$ is a primary producer, $p(S_1^-|\eta_1) \approx 0$ regardless of its niche value $\eta_1$ -- the first niche value among the rank-ordered species -- and this is due to the constant influx of primary productivity $g$, providing $g$ is sufficiently large.
If $g$ is reduced, there is a corresponding increase in $p(S_1^-|\eta_1)$ (see Supplementary Fig. \ref{fig:lowg}).
For niche values -- and by extention trophic levels -- higher than that of the primary producer, a range of $p(S^-|\eta_{i>1})$ is realized (Fig. \ref{fig:probS}A), depending on the particular niche value of the rank-ordered species.
Excluding primary producers, for lower-trophic species with lower $\eta_i$ values, $p(S^-)$ is high when $\eta_i$ is low, declining quickly with higher niche values (Fig. \ref{fig:probS}A).
As one ascends the rank-ordered species to higher niche values along the interval, species have interaction structures that are on average more bottom-heavy, where the number of resources is greater than the number of predators.
These bottom-up forces eventually outweigh top-down effects from predators as the niche interval is ascended, driving down $p(S_i^-)$.
This decline towards lower $p(S_i^-)$ becomes sharper for species further along the rank-ordered set -- corresponding to higher trophic levels -- which are drawing from a larger potential prey base (Fig. \ref{fig:probS}A).
% This result also shows that when a large interval separates a consumer from its potential prey resources along the niche axis (i.e. large $\eta_i$ relative to potential prey), the consumer species is more likely to inhabit a high population state (i.e. low $p(S^-)$). <- short-circuit - discussion material
% [Short Circuiting the food web]?

While the particular niche values of a rank-order set of species has a large influence on state probabilities, the positioning of each species on the niche interval can be constrained by treating the particular placement of each species as a random variable and calculating the expected state probability (top graphic in Fig. \ref{fig:probS}B).
% A food web is generated from the niche model from a set of species with niche values uniformally distributed along the niche interval.
With $\eta_i$ treated as a random variable, the expectation ${\rm E}\{p(S_i^-)\}$ is calculated across a range of possible $\eta_i$ values for the rank-ordered set of species along the niche axis. 
Here, ${\rm E}\{p(S_i^-)\} = \int_{0}^1 p(S_i^-) \beta(\eta_i) {\rm d}\eta_i$, where $\beta(\eta_i)$ is a Beta distribution describing the probabilistic placement of $\eta_i$ on the niche axis.
This distribution has a mean $\mu_i = {\rm E}(\eta_i)$ (where $\eta_i$ values are evenly separated locations along the niche interval), and a variance set as a small proportion of the maximal variance $\sigma_i^2 = \chi\mu_i(1-\mu_i)$, with $\chi = 0.1$.
Once ${\rm E}\{p(S^-)\}$ is calculated, niche values are converted to trophic levels using Eq. \ref{eq:TL}, such that changes in state can be understood in terms of where species fall within the food web (Fig. \ref{fig:probS}B).

The expected state probabilities reveal certain general features, the details of which change with different assumptions of top-down effects.
In all cases, ${\rm E}\{p(S_i^-)\}\approx0$ for primary producers, increasing to a maximum at intermediate trophic levels, and declining again as one nears the top of the food web (gray points in Fig. \ref{fig:probS}B).
The near-zero ${\rm E}\{p(S_i^-)\}$ at low trophic levels is again the consequence of primary productivity $g$ fueling higher probabilities of high population states.
This effect wears off at higher trophic levels.
As one ascends the food web, the ratio of predators relative to prey declines, increasing the role of bottom-up relative to top-down forces, eventually driving ${\rm E}\{p(S_i^-)\}\rightarrow 0$.
Given the predator states deterimining top-down pressure on each species is assumed to be constant (i.e. $p(S_{\rm pred}^-) = c_0$), lowering the assumed value of $c_0$ is observed to increase the maximum $p(S_i^-)$ that is obtained at intermediate trophic levels.
% If $c_0 = 0.5$, species' state probabilities $p(S^-)$ first increase and then decease as one ascends from the bottom to the top of the food web.
% If $c_0 < 0.5$, this increase reaches a higher threshold before descending.
With $c_0 < 0.5$, predators are more likely to exist in high population states, increasing the role of top-down pressures and magnifying ${\rm E}\{p(S_i^-)\}$ across trophic levels (Supplementary Fig. \ref{fig:proSlowpred}).
% Simulation results suggest that - on average - there is a slight increase in $p(S^-)$ along the niche axis, though values do not stray far from 0.5 in either direction. 
% If a slight linear increase in predator state probabilities along the niche interval is instead assumed, rather than the constant value $c_0$, ${\rm E}\{p(S^-)\}$ plateaus at intermediate trophic levels, more closely matching simulation results.
% Integrating a linear increase in $p(S_{\rm pred}^-)$ from 0.45 to 0.55 along the niche axis produces a plateaued ${\rm E}\{p(S^-)\}$ with trophic level, closely matching output from simulations.

That the predicted ${\rm E}\{p(S_i^-)\}$ declining to zero at the highest trophic levels is not observed in simulation results is due to the mismatch in the approximations used for the number of resources ${\rm E}\{n_i\}$ expected for species $i$ at the top of the food web.
% \emph{i} edge effects, where the highest trophic levels have no top-down pressure, and \emph{ii}
Specifically, the approximated number of prey available to species at higher trophic levels is higher than that observed in simulated systems (Supplementary Fig. \ref{fig:predprey}), such that the influence of bottom-up effects will be overestimated in the highest trophic levels of the approximation.
% While the approximation falls off at this this extreme due to the edge effect introduced by the top of the food web, 

% to the expectatins for the number of resources and predators as a function of a species' niche value.
% [Highest trophic level mismatch]

\subsection*{Drivers of trophic cascades across trophic levels}
A trophic cascade describes the top-down impact of a particular predator population on its prey.
A cascade emerges when a predator's high population state drives a low population state in its prey, with effects that may percolate across lower trophic levels.
% and can be evaluated as the relative population size of that species in the presence or absence of a predator's population.
Such a dynamic can effectively be captured by the log response ratio of the prey's population relative to its predator population, where
% \begin{equation}
    $\psi_i = \log(q_1/q_0)$,
% \end{equation}
with $q_1$ and $q_0$ measuring the prey's population in the presence and absence of predator pressure, respectively \citep{fahimipour2017compensation}.
Adapted to the coarse-grained framework, the numerator measures the state of species $i$ when its predator $l$ is in a high population state $S_l^+$, while the denominator measures the state of the species when its predator is in a low population state $S_l^-$.
The cascade effect can then be expressed as
\begin{equation}
    \psi_i = \log \left( \frac{c_1 + (1 - 2p(S_i^- | S_l^+))}{c_1 + (1 - 2p(S_i^- | S_l^-))} \right),
    \label{eq:cascade}
\end{equation}
where the coefficient $c_1$ is in place to avoid infinities, bounding the measure between $\psi_i = \pm \log(c_1 + 1)$.
As the cascade effect $\psi_i\rightarrow +\log(c_1 + 1)$, there is a stronger correlation in the state of species $i$ with its predator, such that when the predator is in a high population state, so is its prey, and this is the antithesis of a trophic cascade.
In contrast, when the effect measure is zero, there is no alignment between the species $i$ and its predator.
As the effect measure $\psi_i\rightarrow -\log(c_1 + 1)$, there is a stronger anti-correlation in the state of species $i$ with its predator, such that when the predator is in a high population state, its prey is in a low population state and vice versa, and this is a hallmark condition of a strong trophic cascade.

Including the approximation of ${\rm E}\{p(S_i^-)\}$ described in the previous section into Eq. \ref{eq:cascade}, and integrating the condition of predator low/high population size, permits a direct evaluation of $\psi$ as a function of species' niche values, and by extension trophic levels.
Because the conditionals in Eq. \ref{eq:cascade} specify changes in predation, they alter only the top down pressure $\Delta_i^{\rm TD}$, which can be expressed
\begin{equation}
    \Delta_i^{\rm TD}|S_l = k^{\rm out}\left[S_l + (2(S-1)C-1)(1-\eta_i)(1-2p(S_{\rm pred}^-))\right],
    \label{eq:condTD}
\end{equation}
where $S_l$ is the state of the predator in question, and $p(S_{\rm pred}^-)$ is the probability of a low population state for the remaining predators.
Eq. \ref{eq:condTD} is then substituted into Eq. \ref{eq:pS} to calculate the conditional probability.

\begin{figure*}[ht!]
    \centering
    \includegraphics[width=0.95\linewidth]{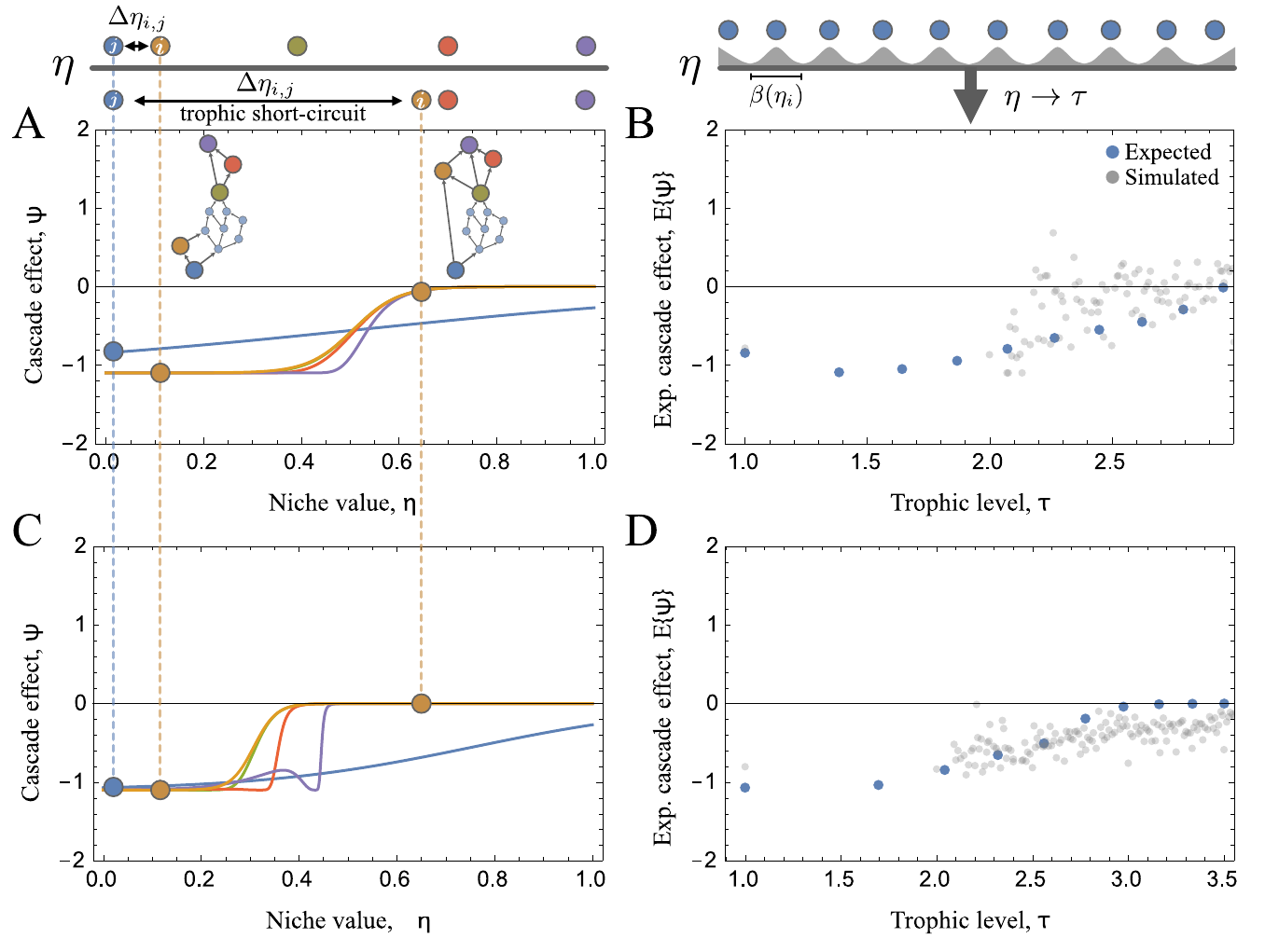}
    \caption{
    The strength of trophic cascades given species richness $S=100$.
    A. The cascade effect for a set of rank-ordered species along the niche interval, as a fucntion of each species $i$'s niche value $\eta_i$, given $C=0.02$.
    B. The expected cascade effect ${\rm E}\{\psi\}$ for a set of rank-ordered species across the niche interval, taken across a small range of potential niche values, given $C=0.02$.
    Niche values have been converted to trophic level using Eq. 2.
    C., D. The same as (A.,B.), assuming $C=0.04$.
    Throughout, we set $c_1 = 2$, meaning that the cascade effect is bounded between $\psi = \pm 1.098$.}
    \label{fig:combinedcascade}
\end{figure*}

As should be expected with the state dynamic imposed in Eq. \ref{eq:statedynamics}, when a predator of species $i$ occupies a high population state, the latter will tend towards a low state, such that $\psi(\eta_i) < 0$, implying that trophic cascades of some strength are to be expected.
While calculation of $\psi(\eta_i)$ shows this to indeed be the case (Fig. \ref{fig:combinedcascade}A), assessement of the cascade effect as a function of species' niche values reveals different strengths of cascades as one moves up the food web.
For primary producers, the cascade effect increases gently with higher placement on the niche axis (colored points in Fig. \ref{fig:combinedcascade}A).
For species with higher placement along the niche axis, and at correspondingly higher trophic levels, $\psi(\eta_i)$ adopts a sigmoidal relationship.

Low-trophic species with higher niche values, while still relying on either primary productivity or lower-trophic prey, are more likely to be consumed by higher-trophic predators, effectively `short-circuiting' the food web.
% Higher-trophic predators consuming very low-trophic prey effectively `short-circuits' the food web.
The presence of a trophic short-circuit is thus captured by a large difference in niche values between species $i$ and its prey $j$, denoted as $\Delta \eta_{i,j}$.
If $\Delta \eta_{i,j}$ is large, much higher-trophic predators, relative to its own trophic level, will be consuming species $i$ (inset, Fig. \ref{fig:combinedcascade}A).
In such a situation, a species $i$ is more likely to be influenced by the presence of the basal resource $g$, while the higher trophic predators are expected to have weaker top-down effects due to the bottom-heavy nature of interactions near the top of the food web. % occupy a high population state given the influence of
Along the short-circuit, these effects combine to reduce the strength of cascades experienced by species $i$, such that $\psi(\eta_i) \rightarrow 0$.
% while at the same time higher trophic predators -- with more resource interactions -- tend to occupy high population states.
% Trophic short-circuits thus serve to reduce cascade effects, resulting in a $\psi(\eta_i)$ closer to zero.
% The below sentence is covered by E(psi) in the next section
% At higher trophic levels, these results suggest that the cascade effect is expected to be diminished.

% In contrast, primary producers consumed by lower trophic predators (corresponding to primary producers with lower $\eta_i$) are anticipated to have stronger cascade effects than those consumed by higher trophic predators (corresponding to primary producers with higher $\eta_i$).
% In these cases,implying diminishing cascade effects at higher trophic levels.
% , which themselves have different bottom-up and top-down pressures.

% For species with higher placement along the niche axis, and at correspondingly higher trophic levels, $\psi(\eta_i)$ adopts a sigmoidal relationship (Fig. \ref{fig:combinedcascade}A).
% Species with placement lower on the niche axis are thus expected to have predators delivering stronger cascade effects, whereas those placed higher on the niche axis experience diminished cascades.
% These species are experiencing the same changes in top-down versus bottom up pressures with varying placement on the niche interval, but without the buffering effects of the basal resource.

If the exact location of species on the niche axis is taken as a random variable (analogous to our calculation of ${\rm E}\{p(S_i^-)\}$), the expected cascade effect ${\rm E}\{\psi_i\}$ is observed to be strongest at lower trophic levels, moving towards zero as one ascends the food web (blue points in Fig. \ref{fig:combinedcascade}B), indicating a diminishment of the cascade strength with trophic level.
Thus, species at higher trophic levels are expected to be less reactive to their predators than those at the bottom of the food web.
These analytical predictions closely align with the lower-boundary of cascade effects calculated from simulated state dynamics under the same conditions (gray points in Fig. \ref{fig:combinedcascade}B).

% The cascade effect $\psi$ increases with trophic level, such that species at higher trophic levels are less reactive to their predators.

Increasing the connectance of the food web alters the predicted cascade effects across trophic levels.
Higher food web connectance increases the number of interactions between species, serving to elevate the number of trophic levels realized within the system.
With connectance doubled from $C=0.02$ to $C=0.04$, the strongest cascades become exaggerated at the lower to intermediate trophic levels, whereas the states of predators and prey at higher trophic levels are increasingly non-correlated (Fig. \ref{fig:combinedcascade}C).
% The potential for cascades to impact primary producers is elevated across the board, whereas the threshold from large to no effect is shifted towards lower niche values for higher-trophic species (Fig. \ref{fig:combinedcascade}C).
Importantly, the realized values of ${\rm E}\{\psi\}$ with respect to a given trophic level -- regardless of connectance -- are roughly the same (Fig. \ref{fig:combinedcascade}B,D).
At the highest trophic levels in these more densely connected food webs, predator-prey interactions are predicted to have little to no correlation, such that ${\rm E}\{\psi\}\rightarrow 0$.
Analytical expectations of the cascade effect ${\rm E}\{\psi\}$ at this higher food web connectance closely matches values generated from simulations (blue versus gray points in Fig. \ref{fig:combinedcascade}D).
It is likely that the mean-field approximation expressed by ${\rm E}\{\psi\}$ is a closer match to simulation results under the assumption of higher connectance due to the averaging effects of more densely-connected species at all trophic levels of the food web.
Importantly, the observed mismatch between analytical solutions and simulations at the top of the food web for $p(S^-)$ (Fig. \ref{fig:probS}B) is not present here because $\psi$ can only be evaluated for species with predators, precluding calculation of ${\rm E}\{\psi_i\}$ for apex predators. 
% In other words, the dynamics between predator and prey at higher trophic levels are less likely to produce strong cascades than those between predators and prey at lower trophic levels.

% If primary producers have a higher likelihood of existing in a high population state due to their reliance on an omnipresent resource, diminishment of $\psi$ with intermidiate $\eta$ implies that higher trophic predators are less likely to be anti-correlated.

% Species with trophic levels higher than those of primary producers reveal a different pattern in $\psi$.
% Here, lower $\eta$ values imply a strong cascade effect, and higher $\eta$ values imply a weak cascade effect where $\psi\rightarrow 0$.

% Higher-trophic predators tend to have a slightly higher potential for existing in a low population state, reducing their ability to suppress their prey.
% If their prey are primary producers 

% The effect of structre on cascades

\begin{figure*}[ht!]
    \centering
    \includegraphics[width=0.95\linewidth]{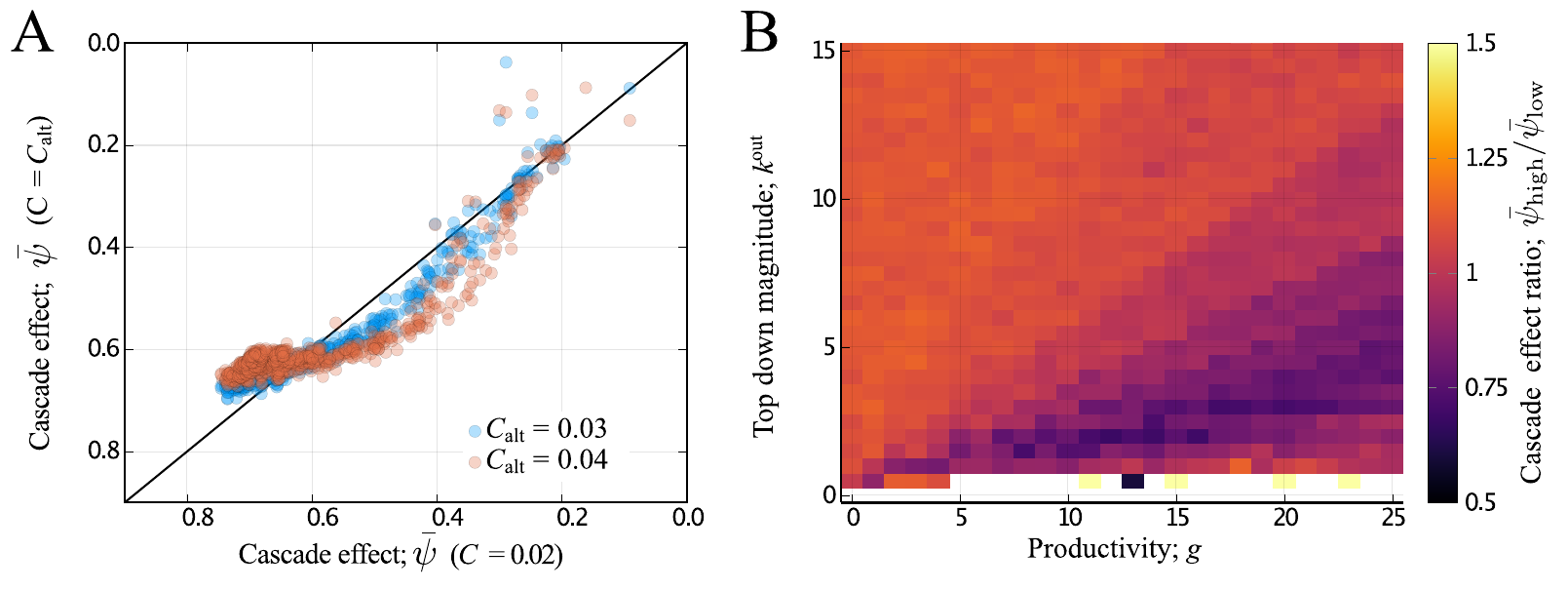}
    \caption{
        Cascade effect as a function of food web connectance.
        A. The average cascade effect $\bar{\psi}$ taken across species in the food web and food web replicates, as a function of primary productivity $g$ and the magnitude of top-down forces $k^{\rm out}$. Values are averaged across 200 replicate food webs, where $k^{\rm in} = 5$, $t_{\rm max} = 300$ (discarding the first 200 time steps) and under the assumption of random initial conditions.
    }
    \label{fig:cascadeconn}
\end{figure*}

While the above exploration of cascade effects as a function of trophic level and connectance is applied to a specific set of dynamic conditions ($g = 10,~k^{\rm out}=10,~k^{\rm in}=5$), different top-down versus bottom-up pressures can lead to different results.
Simulation of the state-dynamic framework across $(g,k^{\rm out})$ reveals that increasing food web connectance can either reduce the effect of trophic cascades (as observed for the specific condition in Fig. \ref{fig:combinedcascade}), or promote them, and this largely depends on the magnitude of top-down effects $k^{\rm out}$ relative to primary productivity $g$ (Fig. \ref{fig:cascadeconn}).
Here, the majority (67\%) of explored $k^{\rm out}$ and $g$ parameter combinations result in an increased $\bar{\psi}$ (where the $\bar{\psi}$ denotes the average cascade effect across entire food webs and across food web replicates) with connectance, indicating a general reduction in the strength of trophic cascades as the density of species interactions increases.
However, systems with elevated productivity (high $g$) and reduced magnitude of top-down effects (low $k^{\rm out}$) reveal the opposite: a decrease in $\bar{\psi}$, meaning a promotion of cascade strength, with increasing connectance (Fig. \ref{fig:cascadeconn}B).
Taken together, it is clear that when bottom-up forces are relatively stronger than top-down forces, higher densities of species interactions are expected to magnify the mean strength of cascades across food webs.

\section*{Discussion}

\begin{quote}
    ``However, not everything that can be counted counts, and not everything that counts can be counted.''
    \\
    -- William Bruce Cameron
\end{quote}

The trophic interactions driving the rise and fall of populations are highly complex, and for many species, unquantifiable.
Where the numbers of individuals within populations are cryptic, uncountably large, or lost to the past, frameworks that are coarse-grained enough to more accurately represent our state of knowledge of such systems, and explore their dynamics, are needed.
A minimal yet meaningful population framework must account for relative differences between interacting populations, where, following both \citet{hanski1985single} and more recently \citet{guimaraes2024dynamics}, a species' population can occupy either a high or low state.
Distributed across an interaction network such as a food web, a set of Boolean rules can then be defined such that clear dynamics determine under what conditions a population changes from a low to a high state, or a high to a low state.
By replacing the complex functional relationships governing species' rate laws with a small set of algorithmic rules determining changes in population state \citep{robeva2016spruce}, one can more readily classify the emergent behaviors, identify stable configurations, and track key transitions in otherwise intractable ecological networks.

% the large number of system states can be catalogued and charted \citep{examples}.
% While such state transition graphs are powerful tools, they are often difficult to explore when networks are large, or the rules determining changes in the states of nodes are complex.

Yet algorithmic rules specifying changes in populations impose their own constraints, and in particular are \emph{i}) difficult to generalize across systems with alternative algorithmic rules \citep{Seshadhri2016}, and \emph{ii}) less open to analytical exploration, particularly when networks are large and complex.
Boolean architectures that are directed towards simplifying population dynamics, yet are algorithmically minimal enough to apply to a wide variety of systems, and can be explored analytically, represent an intermediary between Lotka-Volterra-style frameworks and more complex and ecologically-detailed Boolean approaches \citep[cf.][]{robeva2016spruce}.
Here a general coarse-grained population framework is described that is one such intermediary, where the straightforward balance of top-down versus bottom-up effects determines the change in state of a particular population.
This framework thus maps a Lotka-Volterra-type mechanism to a Boolean rule that can be explored both numerically and analytically.
Leveraging the statistical properties of model food web networks, alongside the single algorithmic rule expressed in Eq. \ref{eq:statedynamics}, expectations of state probabilities and cascade effects can be directly expressed, which is not typically feasible when interactions are more complex.

%Cycles and Duration -- importance of top-down and absence of productivity paradox
\textbf{Dynamics across trophic levels.} Whether predator-prey populations reach a steady state or cycle directly affects their susceptibility to extinction and is therefore of central interest to both theorists \citep{kendall1999populations} and empiricists \citep{myers2018population}.
There are many causes of cyclic behavior in trophic systems, from high growth rates in discrete systems \citep{may1987chaos}, to the adaptation of prey to predator behavior \citep{abrams1997relationship}, to the interaction of consumer starvation timescales with resource availability \citep{yeakel2018dynamics}.
Regardless of the ecological cause, there are some general requirements for cycling populations: that reproduction is sufficiently high, that growth is ultimately regulated by mortality, and that recovery is delayed \citep{myers2018population}. %
% Regardless of the specific ecological cause, there are some general requirements for cycling populations: that reproduction is sufficiently high, a mortality that regulates this growth, and something to prevent immediate recovery \citep{myers2018population}. %Population cycles: generalities, exceptions and remaining mysteries

In continuous time predator-prey models, the potential for cycling can be boiled down to two conditions: \emph{i}) that the contribution of predator-induced mortality is high relative to other sources \citep{yeakel11}, and \emph{ii}) that predator saturation increases with prey availability \citep{gross2004enrichment,stiefs2010food}, regardless of the specific functional forms governing the predator-prey interaction.
As is observed in the results presented here, the proportion of species that cycle tends to increase when $k^{\rm out}$ is high relative to primary production (Fig. \ref{fig:propcycling}).
In the state dynamics defined in Eq. \ref{eq:statedynamics}, top-down predator pressure is captured by $k^{\rm out}$, thus mapping onto criteria \emph{i}), where higher values correspond to predator populations exerting larger negative effects on prey.

On the other hand, the above criteria \emph{ii}) in effect describes the capacity of predator populations to overshoot the ability of their prey resources to support them.
This `overshoot-driven instability' is not possible with a two-state framework.
Along similar lines, the paradox of enrichment emerges from a similar process \citep{rosenzweig1971paradox}, where resource growth fuels consumer growth past its means of support, and this dynamic is also precluded from two-state frameworks.
Instead, as productivity increases, food webs are observed to become static (Fig. \ref{fig:duration}A), where all populations eventually achieve a high-population state.
This stasis is undone only if high productivity is balanced by similarly opposing top-down forcing.
Where bottom-up effects of productivity are very high, and the magnitude of top-down effects is very low, is generally not realized in typical food webs except under highly disturbed or artificial conditions.
In such cases, including ephemeral algal blooms \citep{Dorgham2014} and early-stage successional communities \citep{dunck2021priority}, the destruction or absence of higher-trophic species, at least temporarily, also tends to be observed.

Coarse-graining populations to high and low states permits essential features of trophic systems to be captured with comparatively simple Boolean architecture, without sacrificing analytical utility.
And because the dynamics are minimal, the effects of structure on dynamics can be directly investigated.
From the niche model, it is clear that top-down and bottom-up interaction structures change with trophic level, and two measures -- the probability that species $i$ occupies a low population state $p(S_i^-)$ and its susceptibility to a trophic cascade $\psi_i$ -- can be approximated from state dynamics.
% , though approximation of  is problematic for apex predators, 

Both analytical and simulation results indicate that $p(S_i^-)$ is generally higher at intermediate to high trophic levels, and lower at the base of the food web (Fig. \ref{fig:probS}B).
Assuming that consistent occupation of low population states for long periods of time correlates with extinction risk, this suggests that higher-trophic species are more prone to extinction based entirely on the interaction between food web structure and the competing demands of top-down versus bottom-up forcings.
%  (Fig. \ref{fig:durationtrophic}).
In natural systems, higher-trophic species appear more prone to extinction \citep{purvis2000predicting}, and this is typically attributed to a large number of factors that tend to correlate with trophic level, rather than the structural differences associated with trophic levels.
For example, higher trophic species tend to have larger body size \citep{arim2010food}, lower population densities \citep{Damuth1987}, longer timescales of reproduction \citep{kiltie1982intraspecific}, and are disproportionately targeted or impacted by humans \citep{bradshaw2024small,bradshaw2021relative}.
If there is an inherent dynamical disadvantage for those species occupying the top of the food web -- external to specific traits -- understanding how this may be amplified or diminished by local interactions may be an important next step, and one that a coarse-grained framework is well-suited to tackle.

\textbf{Dynamics of trophic short-circuits.} Approximation of cascade effects $\psi_i$ point to the potential importance of trophic short-circuits within food webs (Fig. \ref{fig:combinedcascade}).
Interactions that short-circuit food webs are those that connect higher-trophic consumers to lower-trophic resources (inset, Fig. \ref{fig:combinedcascade}A), and have received only marginal attention.
For example, short circuits can be created by large-bodied fish consuming lower-trophic crabs in mangrove environments \citep{Sheaves2000}, or apex killer whales relying on lower-trophic mysticetes \citep{springer2003sequential,estes2009causes}. 
These types of cross-cutting interactions appear to be more common in marine systems where food chains are longer, though to what extent their presence impacts dynamics has not been thoroughly investigated.
\citet{Sheaves2000} proposed that trophic short-circuits may reduce predation pressure on higher-trophic alternative prey.
If an apex predator specializes on lower-trophic prey, which may be more abundant or contain a disproportionate amount of biomass \citep[e.g. mysticete whales prior to over-harvesting;][]{roman2003whales}, higher-trophic alternative prey may be off the hook.
However, an opposing argument could be made that such lower-trophic prey subsidize -- and boost the populations of -- apex predators, which may put higher-trophic alternative prey at risk.
This would certainly be a problem if the trophic short-circuit was removed, forcing heavily subsidized apex predators to find food elsewhere.

The approximation of state probabilities (Fig. \ref{fig:probS}) and cascade effects (Fig. \ref{fig:combinedcascade}) presents direct insight into the impact of trophic short-circuits on species.
The niche interval separating species in the niche model, $\Delta \eta_{i,j}$, is a direct proxy for the trophic distance separating species in food webs.
Large trophic distances separating predators and prey effectively represent a trophic short-circuit.
For a given species pair, increasing $\Delta \eta_{i,j}$ captures the presence of a trophic short-circuit (inset, Fig. \ref{fig:combinedcascade}A), which is observed to
\emph{i}) reduce the probability that the predator occupies a low state (Fig. \ref{fig:probS}A), and
\emph{ii}) push the cascade effect $\psi$ towards zero (Fig. \ref{fig:combinedcascade}A), such that the strength of the trophic cascade is lessened.
Together, these results indicate that consumers interacting with trophically-distant prey are more likely to occupy high population states, and because lower-trophic prey are themselves boosted by primary productivity, the high population states of both serves to lessen the strength of cascades.
Short-circuits are thus observed to weaken trophic cascades because apex predator fluctuations do not propagate as strongly downward —- instead, the input of primary productivity dampens fluctuations.

That trophic short-circuits have weakened cascades suggests that these cross-cutting interactions may be less susceptible to disturbances.
The killer whale-mysticete example, in which a large-bodied marine apex predator is thought to have consumed lower-trophic mysticete prey prior to their decline due to industrial whaling, is a historical instance of a potentially ancient interaction motif.
In early Pliocene ocean systems, megatooth sharks such as the extremely large-bodied \emph{Otodus megalodon} at times fed at lower trophic levels compared to the relatively smaller-bodied \emph{Carcharodon} sharks known to specialize on higher-trophic pinnipeds \citep{mccormack2022trophic}.
This has been hypothesized to arise from the reliance of \emph{Otodus} on lower-trophic mysticetes and sirenians, in effect introducing a trophic short-circuit.
In yet older marine communities from the Jurassic, the presence of large-bodied filter-feeding fish \citep{friedman2010} alongside apex marine reptiles appears to set the conditions for similar short-circuits, suggesting that such structures may be important and long-standing components of oceanic communities.

The historical reliance of killer whales on large-bodied mysticete prey set the stage for a more recent regime shift, offering insight into where the framework presented here may be usefully expanded.
As great whale populations declined during the last 300 years of industrial-scale overharvest \citep{baker2004modelling}, some have hypothesized that killer whale populations iteratively switched to smaller-bodied prey such as pinnipeds \citep{springer2003sequential,estes2009causes}, only recently adopting contemporary diets, which in some populations consist largely of sea otters \citep{estes1998killer,williams04}.
While the coinciding effects of anthropogentic pressures on pinniped populations introduces some complexity, the dietary shift of killer whales to pinnipeds may have introduced strong trophic cascades, accelerating their decline, and setting the stage for killer whales switching to smaller prey.
Certainly their reliance on sea otters in contemporary systems has been known to generate strong cascading effects, famously resulting in phase shifts between kelp forests and urchin barrens, depending on the strength of higher trophic interactions \citep{estes1974sea,estes2004complex}. 
% Importantly, the killer whale-mysticete short-circuit, and its likely weakening over time as great whale populations have declined from over-harvest, also showcases a clear limitation of the coarse-grained framework.
% Elimination or lessening of this trophic short circuit may have promoted the emergence of strong trophic cascades between killer whales and alternative prey such as harbor seals, fur seals, sea lions, and sea otters \citep{springer2003sequential,estes2009causes}.

Yet the coarse-grained framework presented here indicates that when prey are predated upon by higher-trophic predators, whether those prey are close to the bottom of the food web (as with mysticetes in the short-circuit) or closer to the top of the food web (as with pinnipeds when killer whales are forced to switch), they are expected to experience, on average, weaker cascades.
%  -- such as killer whales relying on pinnipeds -- are expected to have, on average, weaker cascades ($\psi \rightarrow 0$).
This apparent discrepency points to an interesting phenomenon.
If killer whale populations were held at high population sizes in the presence of the mysticete short-circuit, the elimination of this interaction may have introduced a much stronger top-down impact on the alternative, and higher-trophic pinniped prey.
As with the paradox of enrichment, such overshoot-driven instabilities cannot be captured with a two-state model.
Expansion of the framework presented here -- perhaps merely by adding an additional state -- may provide the necessary ingredients needed to capture this ecological reality.
% the inability of a two-state population model to capture such overshoot-derived destabilization is a clear limitation.

% While transient dynamics such as rapid prey switching and overshoot-driven dynamics suggest the short-term instability that can emerge following major ecological disruptions, the long-term structure of food webs is often shaped by more persistent, stabilizing forces. 
% At steady state, the structure and connectance of food webs play a dominant role in regulating cascade strength.
Transient responses like prey-switching and overshoot-driven instabilities do not occur in isolation; they unfold within food webs that are themselves constrained by broader structural properties. 
As food webs increase in connectance and trophic complexity, these transient dynamics may be buffered by omnivory and interaction redundancy, ultimately influencing the extent to which cascades propagate.
Increasing the connectance of the food web is observed to, on average, increase the mean cascade effect $\bar{\psi}$, which reduces the strength of trophic cascades.
Where top-down effects are very low relative to primary productivity, trophic cascades are instead strengthened (Fig. \ref{fig:cascadeconn}), though it is important to note that most of this parameter region also results in state dynamics that are prone to stasis, and of less ecological relevance.
That most scenarios result in increased $\bar{\psi}$ is driven by the fact that a larger density of interactions results in food webs with more trophic levels, and those higher trophic levels support relatively weaker trophic cascades (Fig. \ref{fig:combinedcascade}B,D).
Empirical food webs with a larger number of trophic levels, such as those in marine environments, tend to have interactions that are more `tangled', with a larger number of omnivorous consumers \citep{thompson2007trophic}.
% With omnivorous interactions, because energy and predation pressure become subdivided across multiple channels, the classical top-down cascade is expected to be diminished.
% Increased prevalence of omnivorous interactions have previously been shown to reduce the likelihood of trophic cascades, as the impacts of interactions are attenuated across a larger number of species spanning multiple trophic levels \citep{bascompte2005interaction}, and this finding aligns with the results presented here.
Increased prevalence of omnivorous interactions has previously been shown to dampen trophic cascades by distributing energy and predation pressure across multiple pathways \citep{bascompte2005interaction}.
The results presented here reinforce this pattern, highlighting how structural complexity inherently stabilizes food webs, reducing the reach of top-down effects in highly connected systems.

\section*{Conclusion}

Distilling higher-dimensional species interactions into simple rules, distributed across complex ecological networks, may offer important insights into the effects of structure on community dynamics \citep{guimaraes2024dynamics}.
By coarse-graining the rise and fall of populations into a Boolean switch that depends on the balance of top-down to bottom-up forcings, basic trophic relationships are captured, while retaining analytical exploration.
An alternative way to analyze population state transitions is through a graph-theoretic or Markovian state-space approach, where all possible system configurations are represented as nodes in a graph, with edges weighted by transition probabilities between states \citep{Froyland2001}. 
Such methods provide a powerful mathematical framework for studying the global connectivity of state transitions, identifying steady-state distributions \citep{pawlowski2009markov,daniel2016state}, and analyzing how populations move through different configurations over time \citep{robeva2016spruce}. 
However, such an approach does not typically encode the ecological mechanisms driving transitions between states, and its computational complexity grows exponentially with the number of species, as the number of possible system configurations scales as $2^S$ for $S$ species.
% such as the shifting balance of top-down and bottom-up forces across trophic levels -— a crucial ingredient for understanding species persistence and the emergence of trophic cascades. 

The framework presented here instead takes a local, mechanistic perspective, directly linking species interactions to state changes in a way that reflects how food webs are structured. 
By coarse-graining populations into high and low states, ecological realism is preserved alongside analytical tractability, allowing an exploration of how food web structure shapes stability, trophic cascades, and the potential importance of trophic short-circuits. 
% While both perspectives offer valuable insights, our approach ensures that theoretical predictions remain firmly rooted in the biological processes governing species persistence and ecosystem dynamics.
% The results explored here show both the utility and limits of such an approach, and highlight the potential importance of trophic short-circuits.
There are many ways in which this approach could be extended while maintaining its relative simplicity, such as including extinction and/or an intermediate population state, thus enabling overshoot-driven dynamics, an important feature of ecological systems currently missing.
Truly knowing the sizes of populations is impractical for most species, and impossible for others such as those that are extremely cryptic or extinct.
Devising frameworks that more accurately capture our limited knowledge of such systems, and to what extent they can provide insight into real-world dynamics, is a worthwhile endeavor that adds a lightweight tool to the theoretical toolbox.

% \textbf{Conservation of species versus interactions.} Connectance...

% \clearpage
\section*{Author Contributions}
JDY designed, implemented, and analyzed the model. JDY wrote the manuscript.

\section*{Acknowledgments}

I would like to thank G. Breed, A. Fahimipour, M. Gaiarsa, P.R. Guimar\~aes Jr., and S. Kim for insightful comments that greatly improved the quality of this manuscript. This project was supported by National Science Foundation grant eMB-2424994 to JDY.
% Irina Birskis-Barros, Jessica Blois, Nathaniel Fox, Paulo Guimar\~aes Jr., and Emily Lindsey for insightful comments and discussions that greatly improved the ideas and concepts that contributed to this manuscript. These ideas benefited greatly from travel funds provided to JDY from the Santa Fe Institute. This project was supported by National Science Foundation grant EAR-1623852 to JDY.

\section*{Data and Code Availability}

Code and data available on the Zenodo Repository: \texttt{TBD}

%Thompson1997
%Tylianakis2010
%Heinen2020

% \bibliographystyle{amnatnat}
% \bibliography{aa_trophic}

\section*{Supplementary Figures}
\FloatBarrier  % Add this right after \section*{Supplementary Figures}

% Restart numbering and add "S" prefix
\setcounter{equation}{0}
\renewcommand{\theequation}{S\arabic{equation}}

\setcounter{figure}{0}
\renewcommand{\thefigure}{S\arabic{figure}}

\setcounter{table}{0}
\renewcommand{\thetable}{S\arabic{table}}

\begin{figure*}[ht!]
    \centering
    \includegraphics[width=0.4\linewidth]{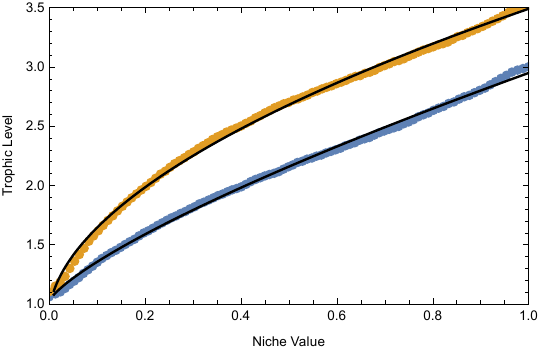}
    \caption{
        Relationship between trophic level and niche value for $C=0.02$ (blue) and $C=0.04$ (orange).
        Black lines represent the best fits, following Eq. \ref{eq:TL}.
    }
    \label{fig:tlniche}
\end{figure*}

\begin{figure*}[ht!]
    \centering
    \includegraphics[width=1\linewidth]{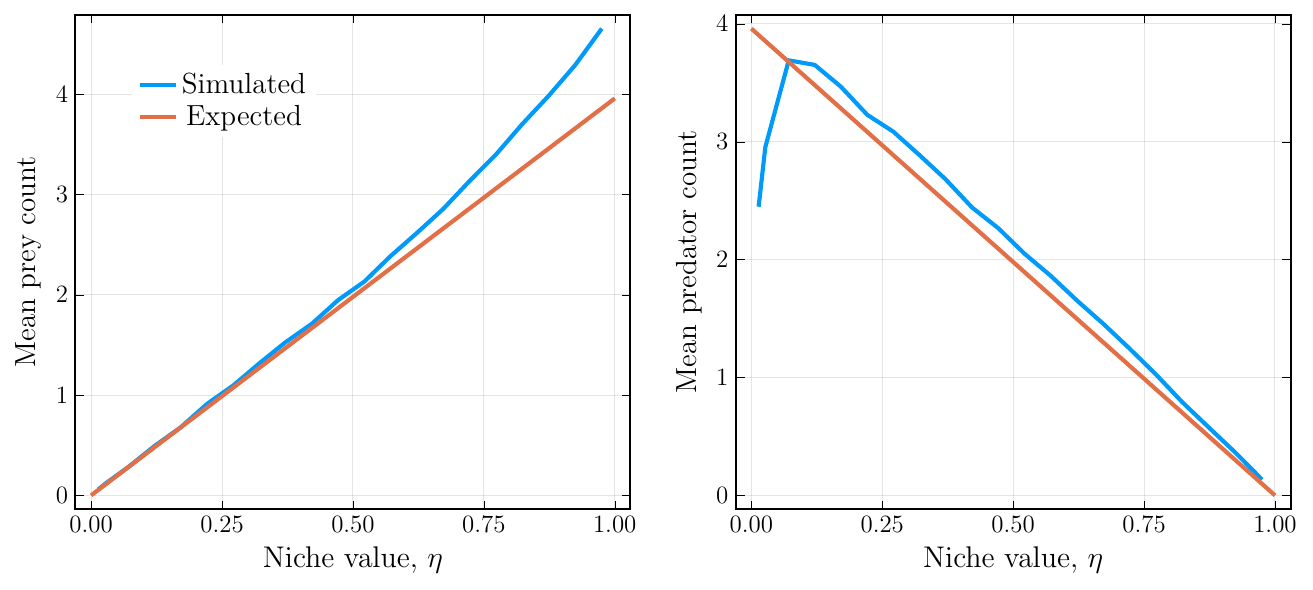}
    \caption{
        A. Simulated (blue line) versus expected (red line) number of prey as a function of a species niche value $\eta_i$ in food webs assembled from the niche model \ref{Williams}.
        The expected number of prey is approximated as ${\rm E}\{n_i\} = 2(S-1)C\eta_i$.
        B. Simulated versus expected number of predators as a function of a species niche value $\eta_i$ in food webs assembled from the niche model \ref{Williams}.
        The expected number of predators is approximated as ${\rm E}\{m_i\} = 2(S-1)C(1 - \eta_i)$.
        Here, $S=100$, $C=0.02$, and mean counts were averaged across 5000 replicate food webs.
    }
    \label{fig:predprey}
\end{figure*}

\begin{figure*}[ht!]
    \centering
    \includegraphics[width=0.6\linewidth]{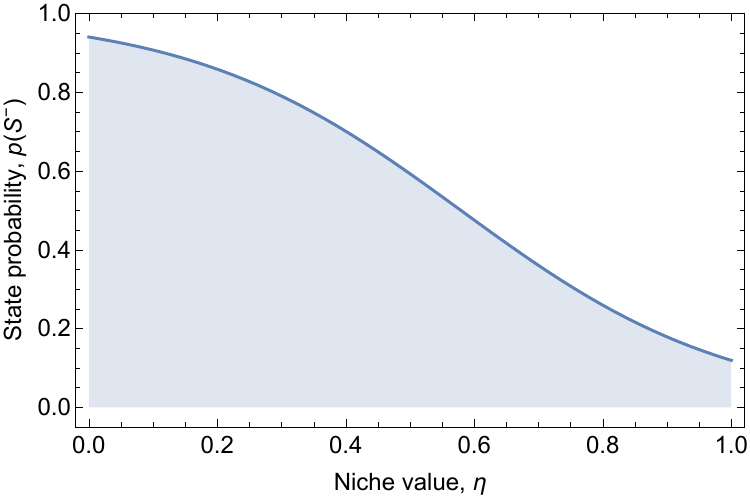}
    \caption{
        The probability of a low population state $p(S^-)$ for a primary producer when productivity is very low ($g=2$) as a function of the primary producer's niche value $\eta_i$. Other parameters are set according to the specific example discussed in the main text, where $S = 100$, $C=0.02$, $k^{\rm out} = 10$, and $k^{\rm in} = 5$.
    }
    \label{fig:lowg}
\end{figure*}

\begin{figure*}[ht!]
    \centering
    \includegraphics[width=0.6\linewidth]{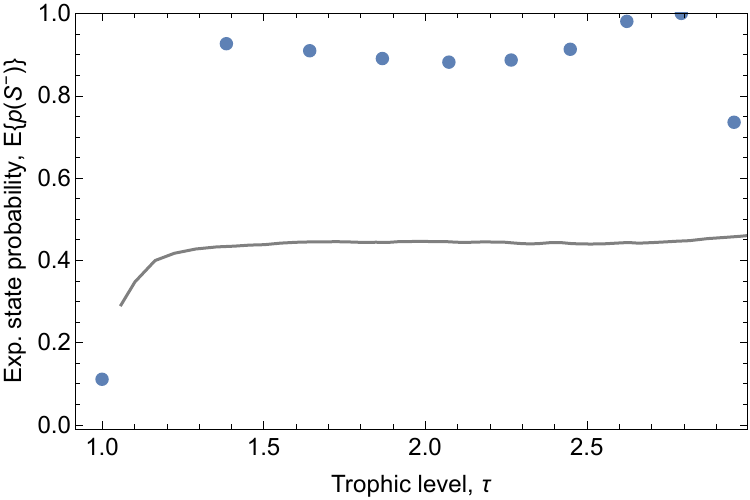}
    \caption{
        The expected probability of a low population state ${\rm E}\{p(S_i^-)\}$ where the probability of predator states $p(S^-_{\rm pred})=0.3$. This means that predator pressures originate from predator populations that are likely to be in high population states, which increases the likelihood that the prey species $i$ is in a low population state. Other parameters are set according to the specific example discussed in the main text, where $S = 100$, $C=0.02$, $k^{\rm out} = 10$, and $k^{\rm in} = 5$.
    }
    \label{fig:proSlowpred}
\end{figure*}

\end{document}